\pdfoutput=1 

\documentclass[11pt]{article}
\usepackage[margin=3.5cm]{geometry}
\usepackage{amsmath,amsfonts,amssymb,amsthm}
\usepackage{times}
\usepackage{framed} 
\usepackage[nottoc,numbib]{tocbibind} 
\usepackage[bookmarks,colorlinks=true,urlcolor=blue]{hyperref}
\usepackage{booktabs}  
\usepackage{color}
\usepackage[affil-sl]{authblk}  
\usepackage{graphicx}
\usepackage{subcaption}
\usepackage{caption} 
\usepackage{verbatim}
\usepackage{url}
\usepackage[normalem]{ulem}
\usepackage[inline, shortlabels]{enumitem}   
\usepackage{setspace} 
\usepackage{tikz}
\usepackage[skins,theorems]{tcolorbox} 
\usepackage[labelformat=parens,labelsep=quad,skip=3pt]{caption}
\usepackage{graphicx}
\usepackage{changepage} 
\usepackage{listings}
\usepackage{cite}  

\newtheorem{rem}{Remark}

\newcommand{\haf}{\ensuremath{\frac{1}{2}}}  	

\def\figs{Figures}    	

\title{Erlang Redux\\[6pt] \Large An Ansatz Method for Solving the M/M/m Queue}
\author{Neil J. Gunther}
\affil{\small Performance Dynamics Company, Castro Valley, CA 94552 \authorcr njgunther@perfdynamics.com}
\date{}

\begin{document}
\maketitle
 \thispagestyle{empty}

\begin{abstract}  
This exposition presents a novel approach to solving an M/M/m queue for the waiting time and the residence time. 
The motivation comes from an algebraic solution for the residence time of the M/M/1 queue. 
The key idea is the introduction of an ansatz transformation, defined in terms of the Erlang B function, that avoids the 
more opaque derivation based on applied probability theory. 
The only prerequisite is an elementary knowledge of the Poisson distribution, which is already necessary for understanding the M/M/1 queue.
The approach described here supersedes our earlier approximate morphing transformation. 
\end{abstract}

\section{Introduction}  \label{sec:intro}
The multi-server M/M/m queue arises in the performance analysis of such systems as: call centers, manufacturing, communications networks, multicore
computers,  and multithreaded software applications.
Unfortunately,  those who should be applying M/M/m models to the performance analysis of their designs and architectures are often not  schooled in 
applied probability theory. This situation cries out for a more intuitive approach to understanding multi-server queues---along the 
lines of the algebraic approach used to develop the residence time for an M/M/1 queue~\cite{qsp,Gunther1998}. 
However, this apparently simple objective has proved more difficult than one might reasonably expect.\footnote{The situation is reminiscent of one that Kepler must have faced in going from circular to elliptic orbits. Introducing even a modest amount of eccentricity causes profound complications for expressing and calculating the circumference of an ellipse. Subsequently, others developed a variety of approximations.}

A previous attempt to meet this goal was based on our {\em morphing model} 
approximation to M/M/m~\cite{Gunther2005,Gunther2017}. The residence time formula in the morphing model  is  simpler mathematically and more
intuitive than the exact solution based on the original Erlang C function~\cite[Eq. 5]{Erlang1917}. 
Nonetheless, it is only an approximation. A similar approach, but one that produces the exact solution, has remained desirable. 

Here, we present a method that achieves the desired goal. 
Our approach arises from a confluence of several observations that had been overlooked previously. 
In particular: 
\begin{enumerate*}[{(i)}]
\item we focus on the mean waiting time $W_m$, rather than the residence time $R_m$  (as was done in the morphing model),    
\item  $W_m$ can be expressed as a transformation of $\mathcal{R}_1$: a fast M/M/1 residence time, 
\item the transformation function $\Phi_B$ takes us from the Erlang B function to the Erlang C function, 
\item since these are probability functions, $\Phi_B$ must exist on the interval $[0,1]$,  and therefore    
\item it cannot be defined in terms of  queue attributes, such as unbounded queue length.
\end{enumerate*}
These observations, taken collectively, then allow us to reprise the logic of the previous morphing derivation to arrive at the exact waiting time 
and residence time formul\ae\	 for an M/M/m queue.

The structure of this paper is as follows.
In Section~\ref{sec:mm1}, we review the algebraic treatment of the M/M/1 queue. 
Section~\ref{sec:morphing} reviews the morphing model, that transforms $m$ parallel M/M/1 queues into a single fast M/M/1 queue, in 
agreement with the residence time characteristics of an M/M/m queue.
The morphing transformation function $\phi_\rho$, which is a finite geometric series in the server utilization $\rho$,  produces only an approximate solution for $R_m$.
Section~\ref{sec:mmm} returns to the original problem but, replaces $\phi_\rho$ with $\Phi_B$ to recover the exact $R_m$. 

 \section{Algebraic M/M/1}  \label{sec:mm1}
The iron law of residence time is 
\begin{equation}
R = S + W  \label{eqn:ironlaw}
\end{equation}
where  $S$ is the mean service time and $W$ the mean waiting time.
The waiting time for M/M/1 can be viewed as being the due to the number of customers in the system, $Q$, 
ahead of you when you join the queue, i.e.,  $W=QS$. 
Furthermore, the number of customers in the system 
can be determined from Little's law, $Q=\lambda R$, where $\lambda$ is the mean arrival rate.

Substituting Little's law into  \eqref{eqn:ironlaw} produces 
\begin{align*}
R	&= S + QS  \\
	&= S + (\lambda R) S \\
	&= S + R (\lambda S) \\
	&= S + R \, \rho  \label{eqn:rtime}
\end{align*}	
where we have denoted the server utilization by $\rho = \lambda S$.
A final rearrangement yields
\begin{equation}	
R_1 = \dfrac{S}{1-\rho}  \label{eqn:mm1R}
\end{equation}
which is the canonical expression for the M/M/1 residence time~\cite{qsp,Gunther1998} but, derived here without resorting to 
 the usual applied probability theory found in standard texts~\cite{Kleinrock,Allen,Robertazzi,Bertsekas,Gross}.
The subscript in \eqref{eqn:mm1R}  has been introduced to distinguish the number of servers, $m$, in the queueing facility for later comparisons.
Notice the restriction $\rho < 1$  in \eqref{eqn:mm1R} to prevent the queue length from becoming infinite (unstable queue). 

\begin{rem}
Although \eqref{eqn:mm1R}---and similar equations that appear throughout---relates mean values of the respective metrics, it is not a so-called 
operational law~\cite{qsp} because these metrics depend on the underlying statistical distribution.  
\end{rem}

\section{Morphing M/M/m}  \label{sec:morphing}
We would like to apply the same algebraic treatment to an M/M/2 queue and ultimately,  its M/M/m generalization,\footnote{It is noteworthy that~\cite{qsp} does not derive or discuss the equivalent of the M/M/m queue.}  especially for more practical applications~\cite{Bertsekas,Gunther2007} and pedagogic purposes~\cite{Classes}.

\begin{rem}
It is important to note that the arrival rate $\lambda$ needs to be doubled for $m=2$ if the capacity of both servers is to be fully utilized. 
Since neither server can be more than 100\% busy, the corresponding server utilization has to be defined as $\rho = \haf \lambda S$ in order
that $\rho < 1$.
\end{rem}

\noindent
Since $\rho^2 <<1$, we expect $R_2 < R_1$ if the denominator in \eqref{eqn:mm1R} is replaced by $1-\rho^2$, viz., 
\begin{equation}
R_2 = \dfrac{S}{1-\rho^2} \label{eqn:mm2R}
\end{equation}
Moreover, we can interpret $\rho^2$ as representing the smaller probability that both servers are busy simultaneously.  
Indeed, \eqref{eqn:mm2R} agrees with the exact solution based on the Erlang's C function~\cite{Erlang1917}. 

Generalizing this observation led to the {\em morphing model}~\cite{Gunther2005,Gunther2017} 
\begin{equation}
R_m(\phi) = \bigg( \dfrac{S}{1-\rho}  \bigg)  \, \phi_\rho    \label{eqn:rmorph}
\end{equation}
where 
\begin{equation}
\phi_\rho =  \dfrac{1-\rho}{1-\rho^m}   \label{eqn:fmorph}
\end{equation}
is the sum of a finite geometric series and 
\begin{equation}
\rho = \dfrac{\lambda S}{m} < 1    \label{eqn:utilC}
\end{equation}
is the per-server utilization.
 
\begin{table}[ht]
\caption{Correction terms for the morphing approximation \eqref{eqn:fmorph}} \label{tab:coefffs}
\begin{tabular}{l|l}
\hline
 \multicolumn{1}{c|}{$\mathbf{m}$} & \multicolumn{1}{c}{Integer polynomials $\mathbf{P_m(\rho)}$ }\\
\hline
1	& $- \rho + 1$ \\
2	& $- \rho^2 + 1$ \\
3	& $3 \rho ^3+\rho ^2-2 \rho -2$ \\
4	& $8 \rho ^4+4 \rho ^3-3 \rho ^2-6 \rho -3$ \\
5	& $125 \rho ^5+75 \rho ^4-20 \rho ^3-84 \rho ^2-72 \rho -24$\\
6	& $-54 \rho ^6-36 \rho ^5+30 \rho ^3+35 \rho ^2+20 \rho +5$\\
7	& $16807 \rho ^7+12005 \rho ^6+2058 \rho ^5-7350 \rho ^4-10920 \rho ^3-8280 \rho ^2-3600 \rho -720$\\
8	& $16384 \rho ^8+12288 \rho ^7+3584 \rho ^6-5376 \rho ^5-10080 \rho ^4-9240 \rho ^3-5355 \rho ^2-1890 \rho -315$\\
\hline
\end{tabular}
\end{table}

\noindent
Equation \eqref{eqn:rmorph} formally captures the idea that an M/M/m queue is load-dependent in such a way that it  
can be regarded as ``morphing'' between two types of virtual queueing facilities: 
\begin{description}
\item[Very low load:] M/M/m acts like a set of $m$ parallel M/M/1 queues with very little waiting-line formation. 
\item[Very heavy load: ] M/M/m  becomes a single M/M/1 queue with a server that is $m$ times faster than a parallel queue server. 
\end{description}

\noindent
According to  \eqref{eqn:rmorph}, adding another server ($m=3$) corresponds to a residence time given by 
\begin{equation*}
R_3(\phi)= \dfrac{S}{1-\rho^3}   \label{eqn:rmorph3}
\end{equation*}
which is incorrect. The exact expression, based on the Erlang C function  \eqref{eqn:myC}, is
\begin{equation}
R_3 = S + \dfrac{3 \rho ^3 \; S}{2 + 2 \rho +  \rho^2 + 3 \rho^3}  \label{eqn:react3}
\end{equation}

The difficulty with \eqref{eqn:react3}, however, is that  it cannot be further simplified, and 
the algebraic form is completely inscrutable by comparison with the morphing model. All intuition is lost. 

Part of the trouble stems from the fact that finite $m$ introduces a truncated exponential series 
and, unlike \eqref{eqn:fmorph} in the morphing model, there is no simple closed-form expression. 
Thus, we are stuck on the horns of a dilemma: the morphing model is much more intuitively appealing (particularly for pedagogy) but it is only an approximation.  
On a beneficial note, although \eqref{eqn:rmorph} is an approximation, the error

\begin{equation}
\Delta R_m(\phi)   <  \dfrac{\ln(m^{1/4})}{1 + \ln(m)}
\end{equation}

\noindent
is bounded above by 25\% for extremely large $m$ values ~\cite{Gunther2017}. In practice, the error is typically  
between 5\% and 10\% and   
that makes the morphing model useful for quick engineering estimates~\cite{Gunther2007}. 
Different approximations for M/M/m queue metrics have been reported by others. See e.g.,~\cite{saka, seidmann}.

 One way out of this dilemma is to find the {\em correction factor} that takes us from  \eqref{eqn:rmorph} to the exact solution. 
 Indeed, the corrected version of  \eqref{eqn:rmorph} can be written as~\cite{Gunther2017}
 \begin{equation}
R_m = \dfrac{S}{1 - \bigg| \dfrac{c_m}{P_{m-1}(\rho)} \bigg| \, \rho^m}   \label{eqn:rcorrectd}
\end{equation}
where $P_{m-1}(\rho)$ is the deflated  polynomial associated with 
\begin{equation*}
P_m(\rho) = c_{m} \, \rho^{m} + \ldots + c_3 \, \rho^3 + c_2 \, \rho^2 + c_1 \, \rho + c_0   \label{eqn:poly}
\end{equation*}
Example integer coefficients, $c_m$,  are shown in Table~\ref{tab:coefffs} for $m=1,2,\ldots 8$.
Clearly, the correction polynomials are just as complicated as the terms in the exact Erlang C function 
so, not much progress has been achieved  by comparison with the morphing model.  

\begin{figure}[!ht]
\centering
\captionsetup{width=0.8\linewidth}
\includegraphics[scale=0.5]{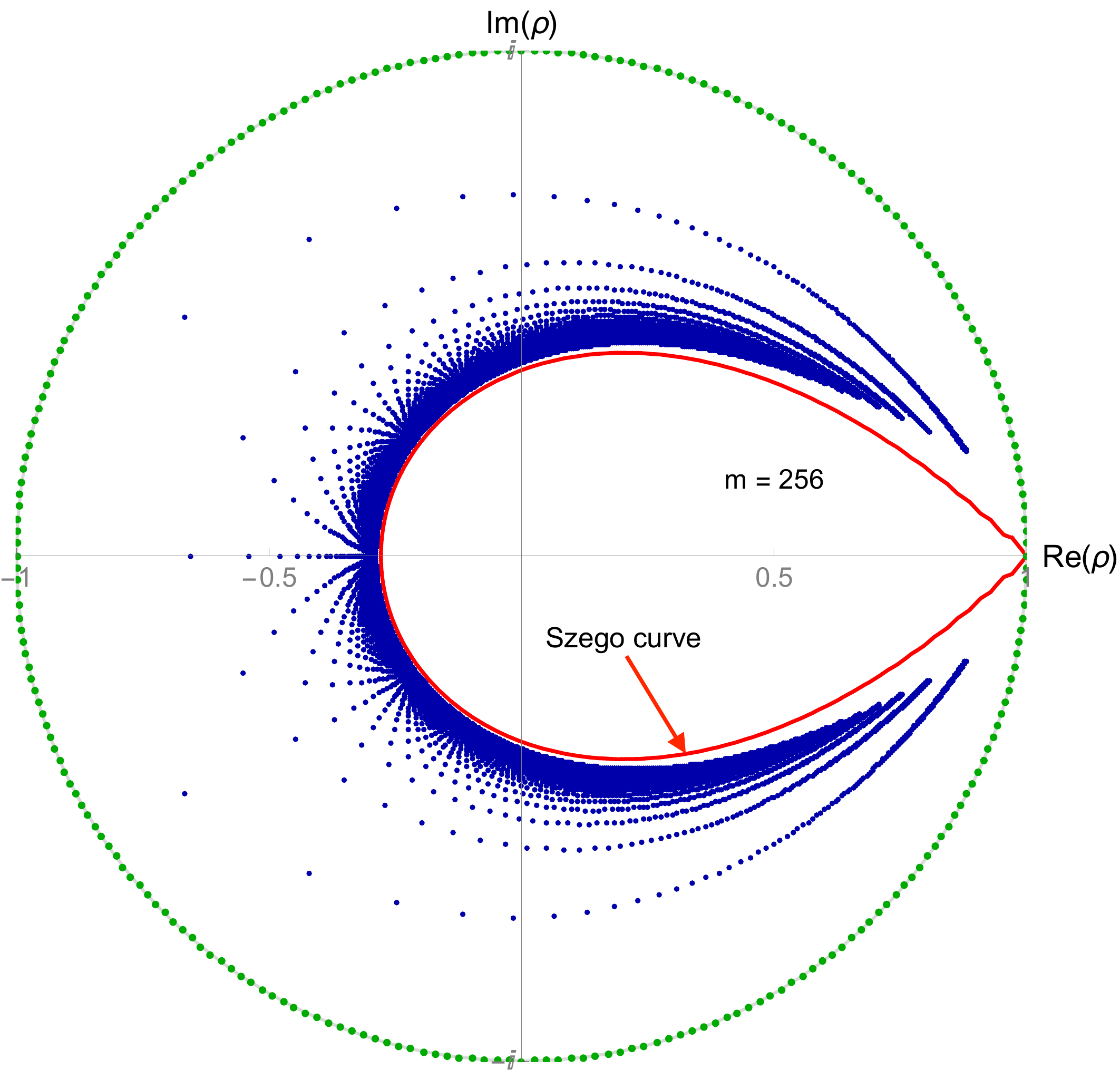}
\caption{Zeros ({\em dots}) of the polynomials in Table~\ref{tab:coefffs} for \mbox{$m=1,2,3,\ldots,256$}. 
Zeros of the morphing approximation ({\em green dots}) lie symmetrically on the circumference of the unit disk. Zeros of the corrected solutions ({\em blue dots}) lie in the interior of the unit disk and converge on the tear-drop shaped Szeg\H{o} bound ({\em red curve}).}   \label{fig:szego}
\end{figure}

The denominator in \eqref{eqn:fmorph},  when analytically continued to complex $\rho$, has zeros that correspond to roots of unity 
that lie on the circumference of the unit disk in Fig.~\ref{fig:szego}. 
Conversely, zeros of the corrected denominator in \eqref{eqn:rcorrectd} lie on the interior of the unit disk. 
As $m$ increases, those zeros move further away from the circumference and converge on the 
 Szeg\H{o} bound~\cite{Gunther2017,Pritsker1997}. 
Even without understanding the mathematical construction, Fig.~\ref{fig:szego} offers a striking visualization of the complexity with which we are dealing.

 \section{Algebraic M/M/m}  \label{sec:mmm}
 Progress toward an algebraic derivation of the exact solution, while at the same time adhering to the objectives of Sections~\ref{sec:intro}   and~\ref{sec:mm1},  can be made by noting that 
the mathematical limitations of the morphing construction (and why it is only an approximation) 
can be attributed to the following assumptions: 
\begin{enumerate}
\item Modifying $R$, rather than $W$, is the wrong starting point.  \label{item:pointW}
\item Unlike M/M/1, both $W_m$ and $R_m$ are state-dependent.
\item The low-traffic limit corresponds to $m$ delay servers, not parallel M/M/1 queues.  \label{item:pointB}
\end{enumerate} 
The last point refers to the assumption that the morphing transformation \eqref{eqn:rmorph}  assumes  $m$ parallel M/M/1 queues,  
with mostly empty waiting lines, in low-traffic limit, whereas there are no waiting states at low load. 

With  assumption~\ref{item:pointW} in mind, we now turn our attention to the canonical exact form of the 
M/M/m waiting time~\cite{Gunther2005,Kleinrock,Allen,Robertazzi} 
\begin{equation}
W_m =  \dfrac{C(m,\rho)  \, S}{m (1-\rho)}   \label{eqn:wm}
\end{equation}
Here, $C(m,\rho)$ is the well-known Erlang C function\footnote{Arnold Allen has described using  \eqref{eqn:myC} to calculate the Erlang C function as  an unnatural act.}, which we write as
\begin{equation}
C(m,a) = \dfrac{A_m}{(1-\rho) \, S_k+ A_m}    \label{eqn:myC}
\end{equation}
with  $a = m\rho$, $A_m = a^m / m!$ and $S_k = \sum_k^{m-1} a^k / k!$.

We want to determine $C(m,\rho)$ by means of a less abstract procedure than that found in either Erlang's original paper~\cite{Erlang1917} or  
standard queueing theory texts~\cite{Kleinrock,Allen,Robertazzi}.   
 The main idea is to reprise the approach used to derive the morphing model but, 
instead of $\phi_\rho$ defined by \eqref{eqn:fmorph}, replace it with 
an ansatz transformation function $\Phi_B(m,\rho)$ to derive the the equivalent of $C(m,\rho)$ in a more intuitive way. 
Once we determine the equivalent of $C(m,\rho)$, the M/M/m waiting time is defined by \eqref{eqn:wm}, and  
the corresponding residence time $R_m$ follows from \eqref{eqn:ironlaw}. 

\def\picsize{0.6}

\begin{figure}[htbp]
\centering
\captionsetup{width=0.8\linewidth}
\includegraphics[scale=\picsize]{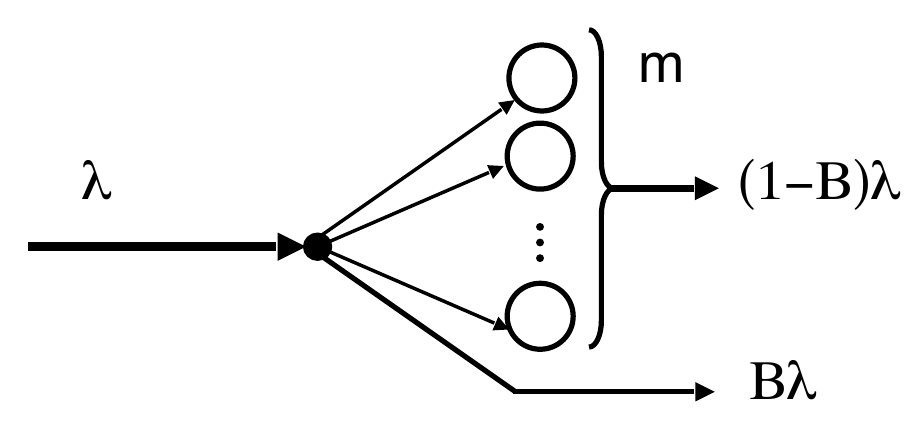}
\caption{A fraction $B\lambda$ of offered calls is rejected and lost from the system when all $m$ servers become instantaneously busy. 
The traffic intensity $a = \lambda S$ can be arbitrarily large.}  \label{fig:qseq1}
\end{figure}

\subsection{Visual development}  \label{sec:qviz}
In this section we adopt the teletraffic parlance of Erlang's paper~\cite{Erlang1917}.
We could start with a pure delay center,  i.e., M/M/$\infty$, where  
calls arrive with mean Poisson rate $\lambda$ and are serviced by an infinite number of servers, each having a mean exponentially-distributed service period $S$. 
Since a call always finds an available operator, no waiting occurs and the mean time spent in the system is simply $R_m=S$.

However, with assumption~\ref{item:pointB} in mind,  it is more appropriate to start with an M/M/m/m queue that has a 
finite number of servers but still no waiting states allowed. 
That restriction causes calls to be lost from the system with probability $B = B(m,\rho)$, as depicted in  Fig.~\ref{fig:qseq1}.
Thus, the queue length can never exceed $m$ calls in service.
This is the Erlang {\em loss model}~\cite{Kleinrock,Allen,Robertazzi,Bertsekas,Gross} with $B$ being  
Erlang's B function~\cite[Eq. 1]{Erlang1917}. Following the notation in \eqref{eqn:myC}, we write it as
\begin{equation}
B(m,a) = \dfrac{A_m}{S_k+ A_m}  \label{eqn:myB}
\end{equation}

\begin{figure}[htbp]
\centering
\captionsetup{width=0.8\linewidth}
\includegraphics[scale=\picsize]{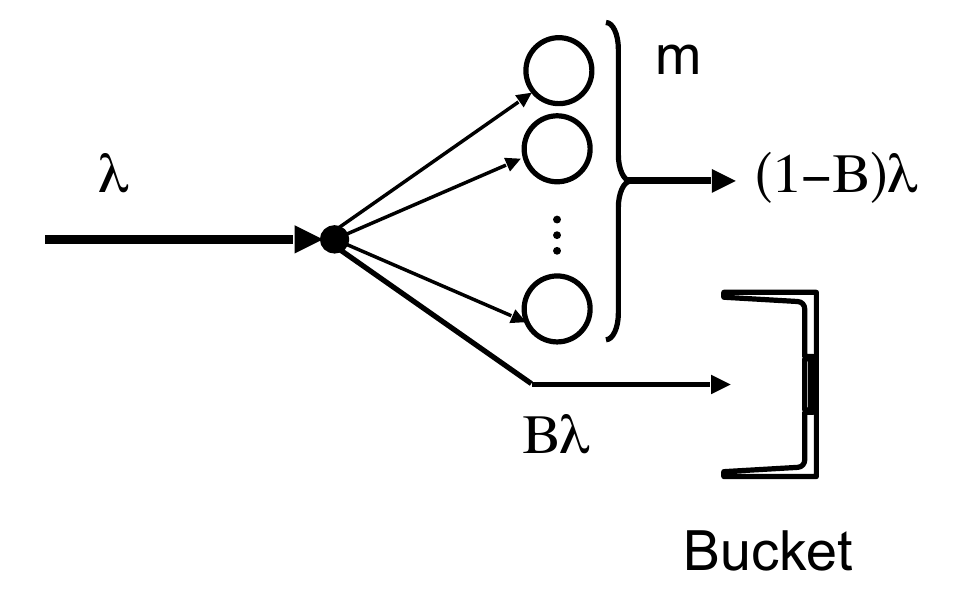} 
\caption{A bucket is introduced to capture the rejected calls as an ordered list. Callers take a number.}  \label{fig:qseq2}
\end{figure}

An M/M/m queue, on the other hand, has waiting states.\footnote{A.K. Erlang called them ``waiting arrangements''  rather than a queue. Callers would presumably wait on the line for the operator to finally  connect their call manually instead of hanging up.} 
In order to include those additional states, we first introduce a ``bucket'' in Fig.~\ref{fig:qseq2} to capture the $B\lambda$ rejected calls.  
These captured calls are placed in an ordered list, i.e., callers take a number.
The bucket does not change the operation of the M/M/m/m queue in any way.
 
Defining
\begin{equation}
\mathcal{R}_1 =  \dfrac{S/m}{1 - \rho} \label{eqn:wmr1}
\end{equation}
to represent the M/M/1 residence time \eqref{eqn:mm1R} but with an $m$-times faster service facility, 
\eqref{eqn:wm}  can be rewritten as
 \begin{equation}
 W_m =  \mathcal{R}_1 \; \Phi_B  \label{eqn:ansatzC}
 \end{equation}
where $\Phi_B$ is a transformation to be determined. 
Equation \eqref{eqn:ansatzC} says that the M/M/m waiting time can be regarded as a proportion 
of the fast residence time $\mathcal{R}_1$. That fraction is given by $\Phi_B$. 
Equation \eqref{eqn:ansatzC} is on the same logical footing as  \eqref{eqn:rmorph} in the morphing model.

Next, the servers in Fig.~\ref{fig:qseq2} are repositioned behind the bucket (with respect to the direction of traffic flow).  
Consequently, the bucket now collects {\em all} incoming calls since there can be no rejected calls. 
This is the first significant differece from Figs.~\ref{fig:qseq1} and~\ref{fig:qseq2}. 
In this configuration, the bucket would accumulate calls indefinitely, due to the fact that none are being serviced, and 
the state-space would therefore become infinite.
 
\begin{figure}[!ht]
\centering
\captionsetup{width=0.8\linewidth}
\includegraphics[scale=\picsize]{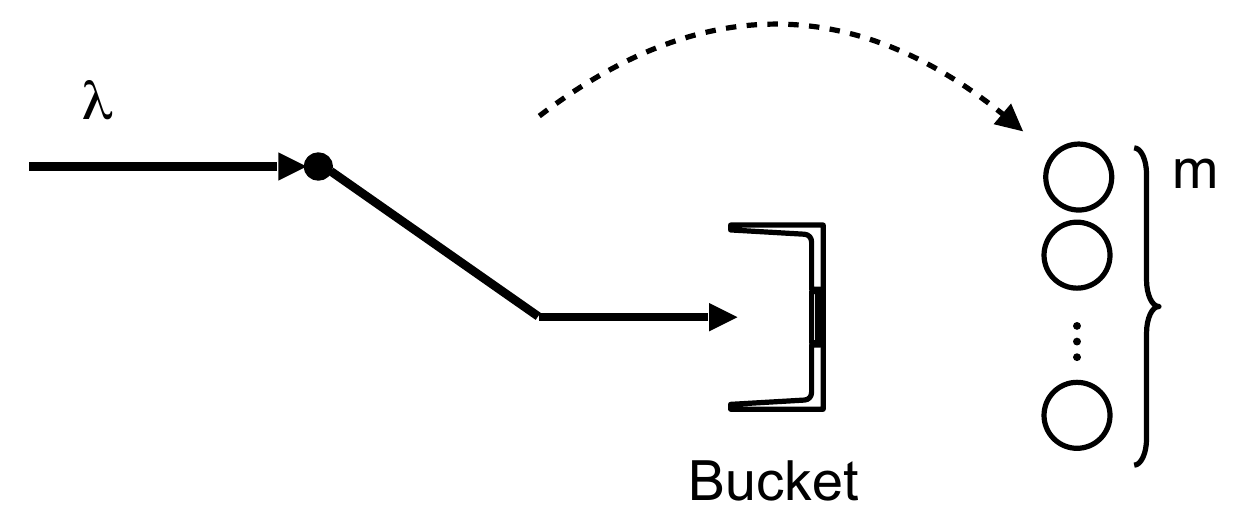} 
 \caption{Next, the servers in Fig.~\ref{fig:qseq2}.are repositioned behind the bucket. 
 The bucket now collects all incoming calls, not just rejected calls, but none are being serviced.}  \label{fig:qseq3}  
\end{figure}

To avoid the ``overflow'' problem in Fig.~\ref{fig:qseq3}, the bucket  has a hole drilled into its base such that calls can be serviced from it in FIFO order.
This is the second significant change. It corresponds to the {\em Erlang C function} in terms of how it relates to the 
Erlang B function.

\begin{figure}[ht]
\centering
\captionsetup{width=0.8\linewidth}
\includegraphics[scale=\picsize]{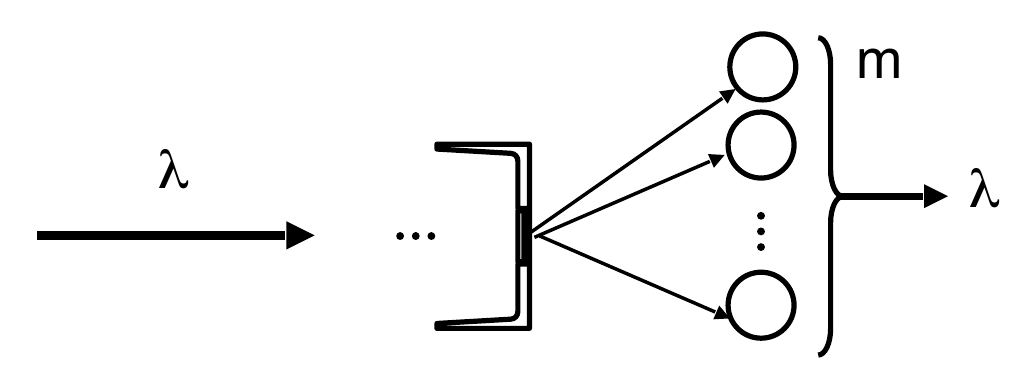} 
 \caption{To service the collected calls, the bucket has a hole drilled into its base such that calls are serviced in FIFO order.
        New arrivals are appended to the ordered list of calls already in the bucket. 
        The traffic intensity is now bounded above by $a  = m$.}  \label{fig:qseq4}  
\end{figure}

Moreover, new arrivals are appended to the ordered list of calls already in the bucket, which  is  equivalent to joining the tail of a waiting line.
With servicing restored, the mean number of requests in the bucket reaches steady-state equilibrium and 
the number of waiting calls becomes bounded.  That number, in turn, determines the mean waiting time $W_m$ in the queue of Fig.~\ref{fig:qseq4}.

\begin{rem}[Utilization]
There is a constraint on the per-server utilization $\rho$ in both an M/M/m/m queue and an M/M/m queue.
Since the effective arrival rate at the M/M/m/m servers, due lost calls in  Fig.~\ref{fig:qseq1}, is only $(1 - B)\lambda$, the per-server utilization is
\begin{equation}
\rho = (1 - B) \, \dfrac{a}{m}  < 1  \label{eqn:rhoB}
\end{equation}
and only approaches 100\% busy at large traffic intensities. 
With the leaking bucket in place (Fig.~\ref{fig:qseq4}), $B = 0$ so, the per-server utilization becomes 
\begin{equation}
\rho =\dfrac{a}{m}  < 1     \label{eqn:rhoC}
\end{equation}
which means that $a < m$, in order to maintain queue stability. 
\end{rem}

\begin{figure}[ht]
\centering
\captionsetup{width=0.8\linewidth}
\includegraphics[scale=0.5]{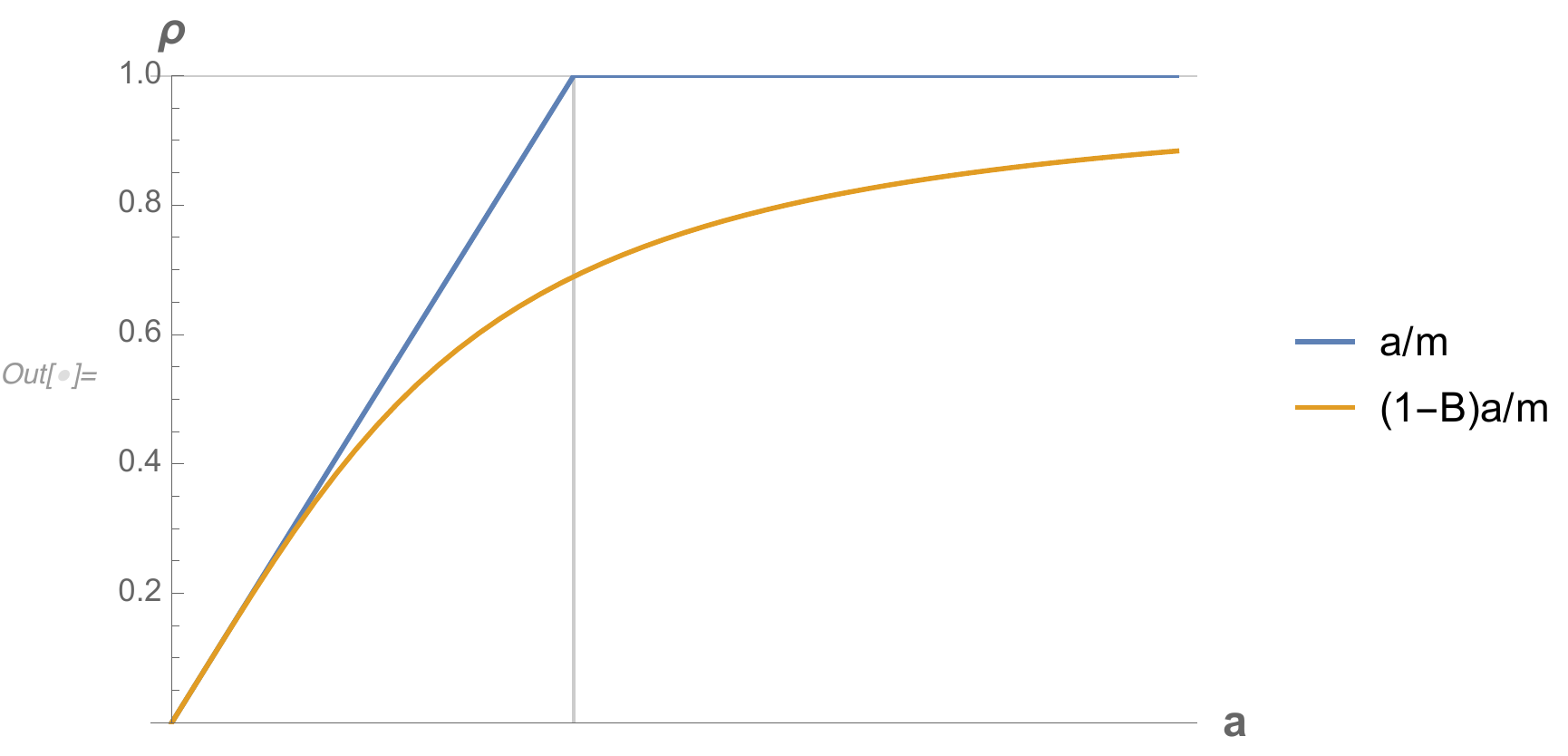} 
 \caption{The per-server utilization in M/M/m/m only approaches 100\% busy as the traffic intensity $a$ becomes very large.  
 M/M/m per-server utilization saturates more rapidly.}  \label{fig:Butil}  
\end{figure}

The difference between \eqref{eqn:rhoB} and \eqref{eqn:rhoC} is shown in Fig.~\ref{fig:Butil}. Arriving calls in 
Figs.~\ref{fig:qseq1} and~\ref{fig:qseq2} are Poisson distributed, and that introduces a tendency toward longer inter-arrival periods, relative to the mean $S$. 
On the other hand, an available M/M/m server instantaneously retrieves the next call from the head of the waiting line 
(the hole in the bucket of Fig.~\ref{fig:qseq4}) and thus, it saturates more rapidly. 

\subsection{Ansatz transformation}   \label{sec:ansatz}
The progression from Fig.~\ref{fig:qseq1} to Fig.~\ref{fig:qseq4} essentially extends the queueing states from a finite state-space in M/M/m/m to 
an infinite  state-space in M/M/m.
We need to include the waiting calls of Fig.~\ref{fig:qseq4} into the transformation function of \eqref{eqn:ansatzC}. 
We know from both M/M/1 and the morphing model that unbounded waiting states are generally identified with the infinite geometric series
\begin{equation}
\dfrac{1}{1 - \rho} = 1 +  \rho +  \rho^2 +  \rho^3 + \ldots  \label{eqn:geoseries}
\end{equation}
familiar in many queue-theoretic formul\ae. 

\begin{figure}[ht]
    \begin{subfigure}[b]{0.5\textwidth}
    	\begin{center}
        \includegraphics[scale=0.45]{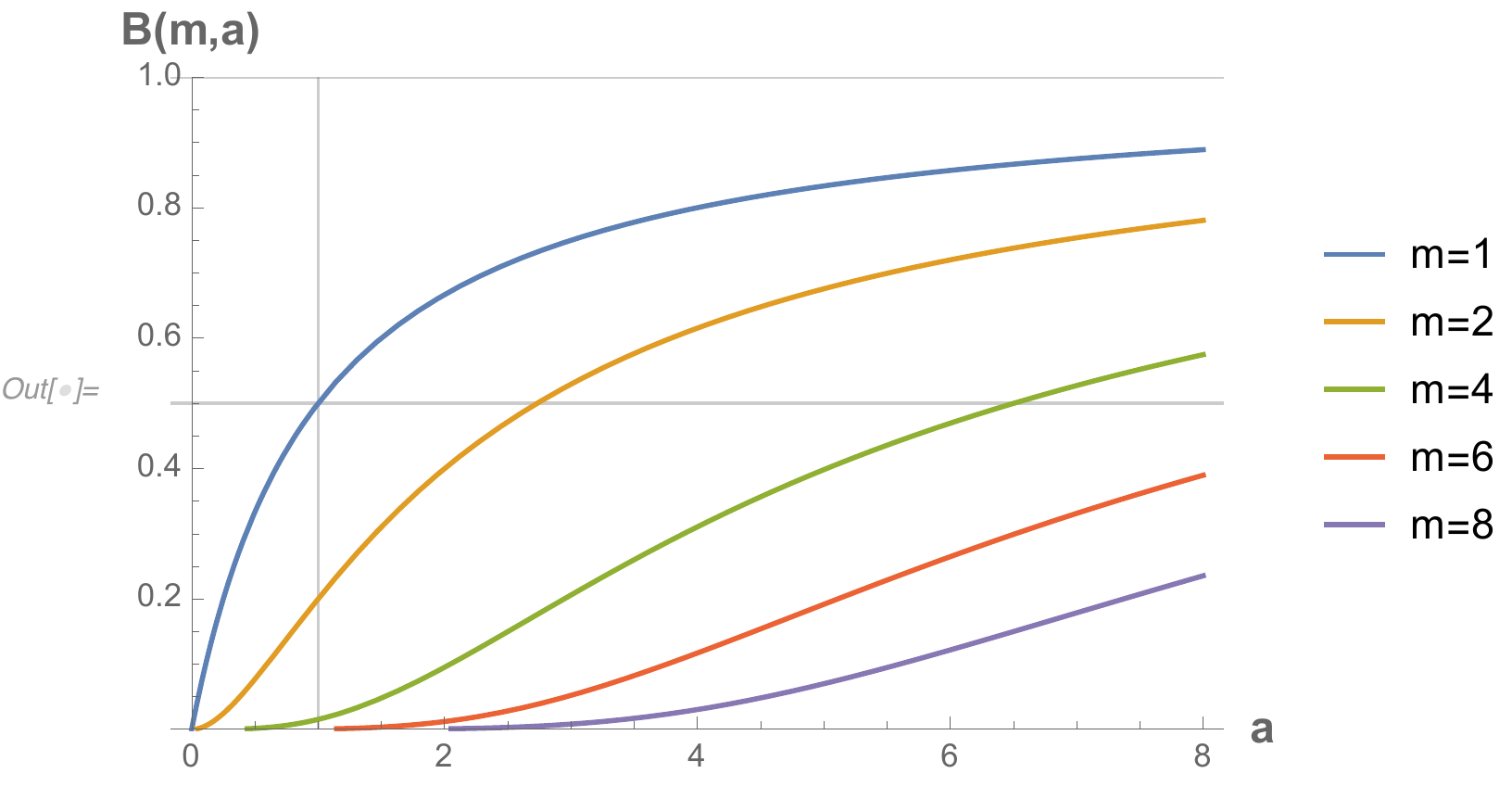}
         \end{center}
        \caption{Blocked call probability, B(m,a)}  \label{fig:EB}
    \end{subfigure}    
    ~ 
    \begin{subfigure}[b]{0.5\textwidth}
    	\begin{center}
        \includegraphics[scale=0.45]{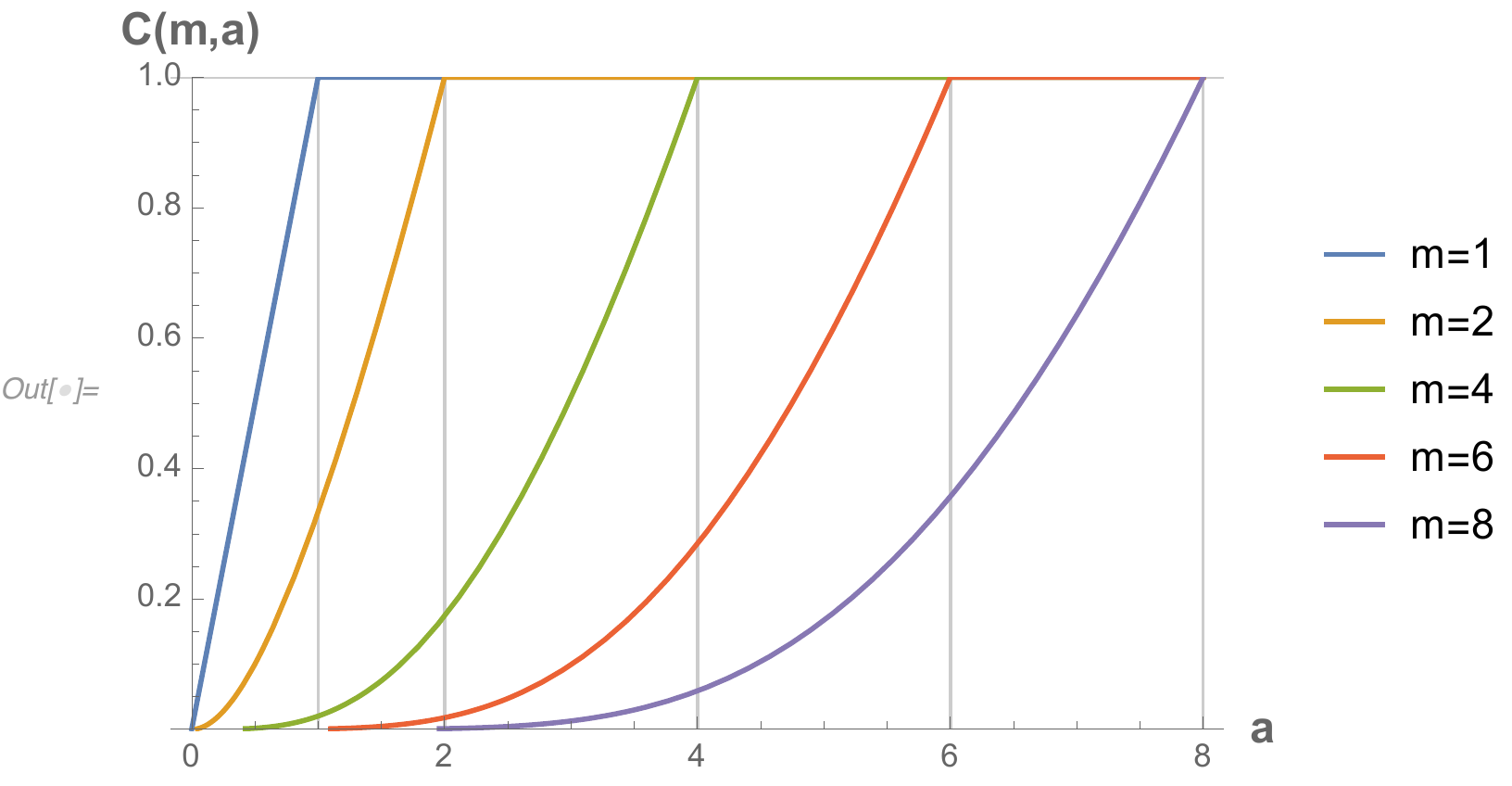} 
        \end{center}
        \caption{Probability that call waits, C(m,a)}  \label{fig:EC}
    \end{subfigure}
    \caption{Erlang B and C curves as functions of the traffic intensity $a=\lambda S$.}  \label{fig:erlangplotz}
\end{figure}

Equation \eqref{eqn:geoseries}  provides a clue as to how we might define $\Phi_B$, starting with $B$ in Fig.~\ref{fig:qseq1} but, 
also including those waiting states.
However, we cannot define $\Phi_B$ in the same way as \eqref{eqn:geoseries}
because Erlang C in \eqref{eqn:myC} is a probability function  that satisfies the following conditions: 
\begin{enumerate}
\item $C(m,a) \in [0,1]; ~\text{for}~m = 1,2,3,\ldots ~\text{and}~ a  \geq 0$.
\item $C(m=1,a)$ is linear-rising in Fig.~\ref{fig:EC}, as expected for M/M/1. For $a > 1$, Erlang C is constant, i.e., \mbox{$C(1,a) = 1$}, since the server remains saturated at 100\% busy. Of course, in this region, an M/M/1 queue becomes unstable. 
\item More generally, $C(m,a)$ is convex  up to $a=m$. 
\item In the low traffic limit $a \rightarrow 0$, we assume $C \simeq B$ (cf. Fig.~\ref{fig:EB}),  
and similarly for our tranformation function, $\Phi_B \simeq B$.
\item In the heavy traffic limit $a \rightarrow m$, we know $C  \rightarrow 1$, which suggests $\Phi_B \rightarrow B/B$.
\end{enumerate}
These considerations lead to the following anzatz for $\Phi_B$: 
\begin{equation}
\Phi_B = \dfrac{B(m,\rho)}{1 - [1 - B(m,\rho)] \, \rho}   \label{eqn:phiB}	
\end{equation}
Example expressions of \eqref{eqn:phiB} are shown in Table~\ref{tab:examples}.

To further substantiate  the choice of \eqref{eqn:phiB}, we consider the light and heavy traffic limits   
\begin{equation}
W_m = \left\{ 
	\begin{array}{ll}
	0 & \text{as} ~\rho = \epsilon \quad \text{(very light traffic)} \\
	\mathcal{R}_1 &\text{as} ~\rho =1-\epsilon  \quad \text{(very heavy traffic)}
	\end{array}
	\right.     \label{eqn:wmlimits}
\end{equation}
where $\epsilon$ is a vanishingly small quantity.

\subsection{Light traffic}
Under very low load, $\rho = \epsilon$, the waiting time \eqref{eqn:ansatzC} becomes
\begin{equation*}
W_m = \dfrac{S/m}{1 - \epsilon}  \bigg[  \dfrac{B(m,\epsilon)}{1 - [1 - B(m,\epsilon)] \epsilon}    \bigg] 
\end{equation*}
Since $B(m,\rho) \simeq 0$ when $\rho = \epsilon$, $W_m$ vanishes.  
Substituting into \eqref{eqn:ironlaw}, the residence time is $R_m = S$, which also corresponds to Fig.~\ref{fig:qseq1} in the low-traffic limit.

\subsection{Heavy traffic}
Under very high load, $\rho = 1-\epsilon$, and \eqref{eqn:ansatzC} becomes
\begin{equation}
W_m = \dfrac{S/m}{1 - (1 - \epsilon)}  \bigg[  \dfrac{B(m,1 - \epsilon)}{1 - [1 - B(m,1 - \epsilon)] (1 - \epsilon)} \bigg]    \label{eqn:wheavy}
\end{equation}

From Fig.~\ref{fig:EB}, we see $B(m,\rho) \ll B(m,a)$ and thus, for a given value of $\rho$ and $m$, $B(m,\rho)$ can be replaced by 
a constant $\delta < 1$. Applying  this to \eqref{eqn:wheavy} produces
\begin{align}
W_m  	&=  \dfrac{S}{m \epsilon}  \bigg[  \dfrac{\delta}{1 - [1 - \delta] (1 - \epsilon)}  \bigg]  \nonumber \\
		&= \dfrac{S}{m \epsilon}  \bigg[  \dfrac{\delta}{1 - (1 - \delta - \epsilon - \delta  \epsilon) }  \bigg]   \nonumber \\
		&= \dfrac{S}{m \epsilon}   \label{eqn:wepsilon}
\end{align}
where we have invoked the additional reasonable assumption $\delta \gg  \epsilon$. Finally,  \eqref{eqn:wepsilon} becomes
\begin{equation*}
W_m =  \dfrac{S}{m(1 - \rho)} = \mathcal{R}_1
\end{equation*}
which is identical to \eqref{eqn:wmr1}, viz.,  an $m$-speed M/M/1 server: 
a result that is also in agreement with the {\em morphing model} of Section~\ref{sec:morphing}. 
As expected, it also corresponds to \eqref{eqn:wm} under heavy traffic 
since Erlang C  reaches  probability one as $\rho$ approaches 100\% busy.

\begin{table}[ht]
\begin{adjustwidth}{-1cm}{} 
\caption{Examples of $\Phi_B(m,a)$ with $a=m\rho$.} \label{tab:examples}
\begin{equation*}
\begin{array}{c | l | l | l}  
\hline 
\mathbf{m} & \multicolumn{1}{c|}{ \mathbf{B(m,a)} } & \multicolumn{1}{c|}{ \mathbf{ [1 - (1 - B(m,a)) \rho]^{-1} } } & \multicolumn{1}{c}{ \mathbf{\Phi_B(m,a)} } \\
\hline 
 1 & \frac{\rho }{1+\rho } & 1+\rho  & \rho  \\[6pt]
 2 & \frac{2 \rho ^2}{1+2 \rho  (1+\rho )} & \frac{1 + 2 \rho+ 2\rho^2}{1+\rho } & \frac{2 \rho ^2}{1+\rho } \\[6pt]
 3 & \frac{9 \rho ^3}{2+3 \rho  (2+3 \rho  (1+\rho ))} & \frac{2 + 6 \rho+ 9 \rho^2 + 9 \rho^3}{2+ \rho  (4+3 \rho )} & \frac{9 \rho ^3}{2+\rho  (4+3 \rho )} \\[12pt]
 4 & \frac{32 \rho ^4}{3+4 \rho  (3+2 \rho  (3+4 \rho  (1+\rho )))} & \frac{3+4 \rho  (3+2 \rho  (3+4 \rho  (1+\rho )))}{3+\rho  (9+4 \rho  (3+2 \rho ))} & \frac{32 \rho ^4}{3+\rho  (9+4 \rho  (3+2 \rho ))} \\[12pt]
 5 & \frac{625 \rho ^5}{24+5 \rho  (24+5 \rho  (12+5 \rho  (4+5 \rho  (1+\rho ))))} & \frac{24+5 \rho  (24+5 \rho  (12+5 \rho  (4+5 \rho  (1+\rho ))))}{24+\rho  (96+5 \rho  (36+5 \rho  (8+5 \rho )))} & \frac{625 \rho ^5}{24+\rho  (96+5 \rho  (36+5 \rho  (8+5 \rho )))} \\[12pt]
 6 & \frac{324 \rho ^6}{5+6 \rho  (5+3 \rho  (5+\rho  (10+3 \rho  (5+6 \rho  (1+\rho )))))} & \frac{5+6 \rho  (5+3 \rho  (5+\rho  (10+3 \rho  (5+6 \rho  (1+\rho )))))}{5+\rho  (25+6 \rho  (10+3 \rho  (5+\rho  (5+3 \rho ))))} & \frac{324 \rho ^6}{5+\rho  (25+6 \rho  (10+3 \rho  (5+\rho  (5+3 \rho ))))}\\[12pt]
\hline 
\end{array}
\end{equation*}
\end{adjustwidth}
\end{table}

\section{Numerics}
Our purpose here has been to offer a more intuitive derivaton of M/M/m queueing metrics, not to promote \eqref{eqn:phiB}  as a  
computational device. Computing  \eqref{eqn:phiB} is equivalent to computing \eqref{eqn:myC}. 
However, if one should want to use $\Phi_B(m,a)$ for calculations or other instruction, then 
it is clear that $B(m,a)$  has to be evaluated first. 

Rather than using \eqref{eqn:myB} which, to paraphrase Arnold Allen: is hardly more ``natural'' than  \eqref{eqn:myC}, 
Erlang B can more easily be computed using the iterative algorithm~\cite{pdsc} in listing~\ref{list:eB}. 

\begin{lstlisting}[language=R, basicstyle=\footnotesize, columns=flexible, keepspaces=true, caption={R code to compute the Erlang B function}, label=list:eB]
erlangB <- function(m, a) {
    eB <- a / (1 + a)
    if(m == 1) { return(eB) } 
    for(k in 2:m) {
       eB <- eB * a / (a * eB + k)
    }
    return(eB)
}
\end{lstlisting}

If R, or similar statistical software, is already being employed, 
one can make direct use of the Poisson PMF (probability mass function) and CDF (cumulative distribution function)  to simplify the code in listing~\ref{list:eB2}.

\begin{lstlisting}[language=R, basicstyle=\footnotesize, columns=flexible, keepspaces=true, caption={Compute Erlang B from the Poisson PMF and CDF}, label=list:eB2]
erlangB <- function(m, a) {
  return(dpois(m, a) / ppois(m, a))
}
\end{lstlisting}

Example calculations computed in this way are summarized in Table~\ref{tab:numerics}.

\begin{table}[htp]
\caption{Example M/M/m metrics  with mean service time  $S=1$~\cite{Erlang1917}}    \label{tab:numerics}
\small 
\begin{adjustwidth}{-0.75cm}{} 
\begin{tabular}{rrlllllll}
\hline
 m  &   \multicolumn{1}{c}{a} &   \multicolumn{1}{c}{$B(m,a)$}  &  \multicolumn{1}{c}{Poisson}  & \multicolumn{1}{c}{$\Phi_B$}  &  
 \multicolumn{1}{c}{$C(m,a)$}  &   \multicolumn{1}{c}{$W_m$} &  \multicolumn{1}{c}{$R_m$}  & \multicolumn{1}{c}{$R_m(\phi)$}\\
 \hline
1  &   0.75 &   0.42857143 &   0.42857143 &   0.75000000 &   0.75000000 &   3.00000000 &   4.000000 &   4.000000\\
2  &   1.50 &   0.31034483 &   0.31034483 &   0.64285714 &   0.64285714 &   1.28571429 &   2.285714 &   2.285714\\
3  &   2.25 &   0.24720244 &   0.24720244 &   0.56775701 &   0.56775701 &   0.75700935 &   1.757009 &   1.729730\\
4  &   3.00 &   0.20610687 &   0.20610687 &   0.50943396 &   0.50943396 &   0.50943396 &   1.509434 &   1.462857\\
8  &   6.00 &   0.12187578 &   0.12187578 &   0.35698109 &   0.35698109 &   0.17849054 &   1.178491 &   1.111251\\
16 & 12.00 &  0.06041259  &   0.06041259 &   0.20457386 &   0.20457386 &   0.05114346 &   1.051143 &   1.010124\\
32 & 24.00 &  0.02209487  &   0.02209487 &   0.08288545 &   0.08288545 &   0.01036068 &   1.010361 &   1.000100\\
\hline
\end{tabular}
\end{adjustwidth}
\end{table}%

\section{Conclusion}
The goal of algebraically deriving the exact residence time for an M/M/m queue---motivated by the same approach to 
M/M/1---has finally been achieved here. 
Several subtle observations are needed to enable this result: 
\begin{enumerate*}[(a)]
\item focus on the waiting time $W_m$, rather than the residence time  $R_m$, 
\item make M/M/m/m the starting point (rather than parallel M/M/1 queues), 
\item the diagrams in Figs.~\ref{fig:qseq1}--\ref{fig:qseq4} aid development of the ansatz $\Phi_B$, 
\item $\Phi_B$ must conform to a probability function, and
\item equation \eqref{eqn:phiB}  modifies the fast residence time $\mathcal{R}_1$, not $R_1$ 
\end{enumerate*} 
These observations also facilitated reprising  the morphing model derivation to verify our ansatz.

Equation \eqref{eqn:phiB}   can also be derived formally from \eqref{eqn:myC} and  \eqref{eqn:myB} but, 
their respective starting points rely on conventional applied probability theory methods, which it has been our objective to avoid.   
Indeed, the same expression for the Erlang C function is known in the literature~\cite{Allen,Robertazzi}, especially for the purpose of programmatic 
computation.



\end{document}